\documentclass[twocolumn,prb,showpacs,floatfix,amsfonts,superscriptaddress]{revtex4-1} 

\usepackage[]{graphicx} 
\usepackage{amsbsy} 
\usepackage{amssymb}
\usepackage[x11names]{xcolor} 			

\newcommand{\sesV}{V$_2$O$_3$}
\newcommand{\sesX}{(V$_{1-x}$Cr$_x$)$_2$O$_3$}
\newcommand{\sesC}{(V$_{0.989}$Cr$_{0.011}$)$_2$O$_3$}
\newcommand{\ag}{$a_{1g}$}
\newcommand{\tg}{$t_{2g}$}
\newcommand{\eg}{$e_g^\pi$}
\newcommand{\esig}{e$^\sigma_g$}

\begin{document}


\title{Evolution of the electronic structure of a Mott system across its phase diagram: an X-ray absorption spectroscopy study of \sesX} 

\author{F.~Rodolakis} 
\affiliation{Laboratoire de Physique des Solides, CNRS-UMR 8502, Universit\'{e} Paris-Sud, F-91405 Orsay, France} 
\affiliation{Synchrotron SOLEIL, L'Orme des Merisiers, Saint-Aubin, BP~48, 91192 Gif-sur-Yvette Cedex, France} 
\altaffiliation{Current address: Material Science Division, Argonne National Laboratory, Argonne, Illinois 60439, USA} 

\author{J.-P.~Rueff} 
\affiliation{Synchrotron SOLEIL, L'Orme des Merisiers, Saint-Aubin, BP~48, 91192 Gif-sur-Yvette Cedex, France} 
\affiliation{Laboratoire de Chimie Physique--Mati\`ere et Rayonnement, CNRS-UMR~7614, Universit\'e Pierre et Marie Curie, F-75005 Paris, France} 

\author{M.~Sikora} 
\affiliation{ESRF, 6 rue Jules Horowitz, BP~220, 38043 Grenoble Cedex, France}  
\affiliation{AGH University of Science and Technology, Krakow, Poland}

\author{I.~Alliot} 
\affiliation{ESRF, 6 rue Jules Horowitz, BP~220, 38043 Grenoble Cedex, France} 
\affiliation{CEA/DSM/INAC/NRS 17 avenue des Martyrs, 38000 Grenoble, France}

\author{J.-P.~Iti\'{e}} 
\affiliation{Synchrotron SOLEIL, L'Orme des Merisiers, Saint-Aubin, BP~48, 91192 Gif-sur-Yvette Cedex, France} 

\author{F.~Baudelet} 
\affiliation{Synchrotron SOLEIL, L'Orme des Merisiers, Saint-Aubin, BP~48, 91192 Gif-sur-Yvette Cedex, France} 

\author{S.~Ravy} 
\affiliation{Synchrotron SOLEIL, L'Orme des Merisiers, Saint-Aubin, BP~48, 91192 Gif-sur-Yvette Cedex, France} 

\author{P.~Wzietek} 
\affiliation{Laboratoire de Physique des Solides, CNRS-UMR 8502, Universit\'{e} Paris-Sud, F-91405 Orsay, France} 

\author{P.~Hansmann}  
\affiliation{Institut for Solid State Physics, Vienna University of Technology, 1040 Vienna, Austria}  

\author{A.~Toschi} 
\affiliation{Institut for Solid State Physics, Vienna University of Technology, 1040 Vienna, Austria}  

\author{M.W.~Haverkort}  
\affiliation{Max-Planck-Institut f\"ur Festk\"orperforschung, Heisenbergstrasse 1, D-70569 Stuttgart, Germany}  

\author{G.~Sangiovanni}  
\affiliation{Institut for Solid State Physics, Vienna University of Technology, 1040 Vienna, Austria}  

\author{K.~Held}  
\affiliation{Institut for Solid State Physics, Vienna University of Technology, 1040 Vienna, Austria}  

\author{P.~Metcalf} 
\affiliation{Department of Chemistry, Purdue University, West Lafayette, Indiana 47907, USA} 

\author{M.~Marsi} 
\affiliation{Laboratoire de Physique des Solides, CNRS-UMR 8502, Universit\'{e} Paris-Sud, F-91405 Orsay, France} 
 
\date{\today}

\begin{abstract} 
\sesV\ is an archetypal system for the study of correlation induced, Mott-Hubbard metal-insulator transitions. Despite decades of extensive investigations, the accurate description of its electronic properties remains an open problem in the physics of strongly correlated materials, also because of the lack of detailed experimental data on its electronic structure over the whole phase diagram. We present here a high resolution X-ray absorption spectroscopy study at the V K-edge of \sesX\ to probe its electronic structure as a function of temperature, doping and pressure, providing an accurate picture of the electronic changes over the whole phase diagram. We also discuss the relevance of the parallel evolution of the lattice parameteres, determined with X-ray diffraction. This allows us to draw two conclusions of general interest: first, the transition under pressure presents peculiar properties, related to a more continuous evolution of the lattice and electronic structure; second, the lattice mismatch is a good parameter describing the strength of the first order transition, and is consequently related to the tendency of the system towards the coexistence of different phases. 
Our results show that the evolution of the electronic structure while approaching a phase transition, and not only while crossing it, is also a key element to unveil the underlying physical mechanisms of Mott materials . 
\end{abstract}


\maketitle


\section{Introduction}
\label{intro}

A full understanding of electronic phase transitions in transition metal oxides still poses many challenges in modern condensed matter physics. In fact, the relevant electronic correlations of partially filled $3d/4d$ orbitals often drive these systems into situations where several energy scales are competing, so that their low energy physics results dramatically affected by very slight changes of the external control parameters like temperature, pressure, doping, etc..  In this class of materials, chromium doped vanadium sesquioxide \sesX\ has attracted considerable interest for undergoing a well known first order correlation driven metal-insulator transition (MIT) between a paramagnetic insulating (PI) and a paramagnetic metallic (PM) phase \cite{McWhan1970}.

\begin{figure}[!tb]
\includegraphics[width=0.85\linewidth]{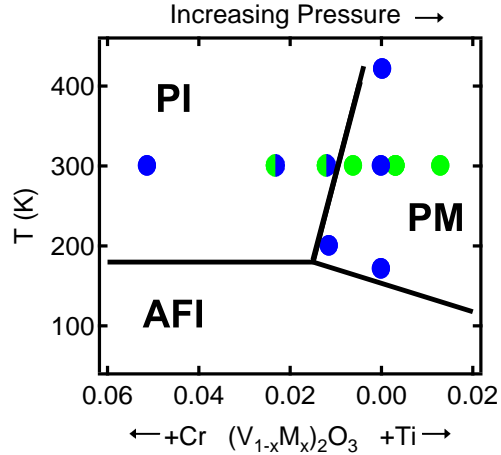}
\caption{(Color online) Phase diagram of \sesX\ as a function of temperature, doping and pressure according to ref.~\cite{McWhan1973}, in which a pressure-doping equivalence has been empirically established to $\pm4$~kbar per percent of substituted atom. The blue (dark gray) dots and green (light gray) dots indicated, respectively, the different temperature, pressure and doping levels studied in this paper.} 
\label{fig:diagram}
\end{figure}

The phase diagram of \sesX\ based on conductivity measurements\cite{McWhan1973} is shown in Fig.~\ref{fig:diagram}. At room conditions, \sesV\ is a paramagnetic metal (PM) with a corundum crystal structure. When the temperature is cooled down below the N\'{e}el temperature it undergoes a structural and electronic transition towards a monoclinic, anti-ferromagnetic insulating phase (AFI), while doping with Cr at room temperature changes it to a paramagnetic insulator (PI), mantaining the original corundum structure. The application of an external pressure has the opposite effect, driving the PI Cr-doped \sesV\ back into the PM phase. For a small amount of chromium doping ($0.9\%<x<1.5\%$), the metallic phase can be  reached by slightly decreasing the temperature. Thus, the isostructural PM-PI transition can be induced by changing either temperature, doping or pressure.  
This transition results in a jump in resistivity of several orders of magnitude\cite{McWhan1969,Kuwamoto1980} and is accompanied by an anisotropic discontinuity of the lattice parameters\cite{McWhan1970}.

\begin{figure}[!tb]
\includegraphics[width=0.95\linewidth]{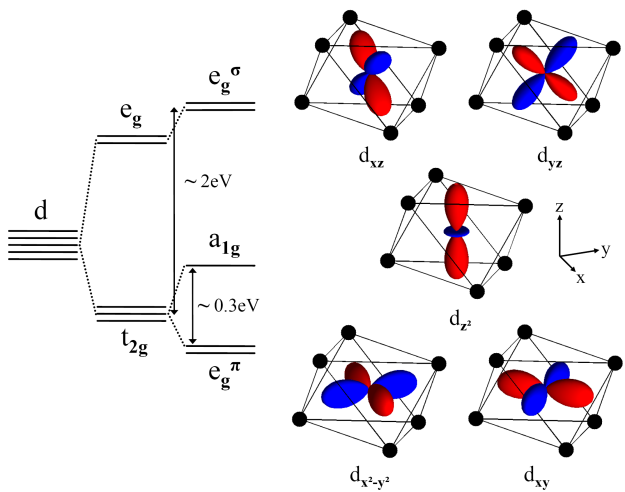}
\caption{(Color online) Schematic representation of the $d$ energy levels with the corresponding orbitals in the corundum structure. Due to the trigonal crystal field, the fivefold degenerate $d$ band splits into three bands: \esig\ ($d_{xz}$,$d_{yz}$), \ag\ ($d_{z^2}$) et \eg\ ($d_{x^2-y^2}$,$d_{xy}$).  Numerical values are obtained from LDA calculation in Ref~\cite{Keller2004}.}
\label{fig:orbitales}
\end{figure}

In the corundum structure, each vanadium atom is surrounded by an octahedron of oxygen atoms, leading to a crystal field splitting of the five-fold degenerate $d$ band into \tg\ and $e_{g}$ bands.
Due to the trigonal distortion of the metallic site, the \tg\ orbitals further split into \eg\ and \ag\ bands, as illustrated in Fig.~\ref{fig:orbitales}.  It is nowadays widely accepted that the $V\ 3d^{2}$ ion has a $S=1$ spin state in all the phases, with a mixed  ($e_{g}^{\pi}$,$a_{1g}$) and ($e_{g}^{\pi}$,$e_{g}^{\pi}$) orbital occupation\cite{Ezhov1999,Paolasini1999}. A combination of V $L_{2,3}$  polarized X-ray absorption measurements and multiplet calculations revealed a large redistribution in the orbital occupation at the transition, with a ratio  $(e_{g}^{\pi}$,$a_{1g}:e_{g}^{\pi}$,$e_{g}^{\pi})\approx1:1$ and $3:2$ in the PM and PI phase, respectively\cite{Park2000}. Taking the multi-orbital nature of the MIT into account, recent theoretical models based on LDA+DMFT calculations succeeded in reproducing the main experimental features of the doping induced PM-PI transition\cite{Keller2004,Poteryaev2007}.

Experimentally, one of the most challenging issues is to analyze on equal footing the strong changes occurring in the different regions of the phase diagram of this compound under the effect of different external parameters.

In order to help clarify this important issue, we present a high resolution X-ray absorption spectroscopy (XAS) study at the V $K$ edge of \sesX\ in the two paramagnetic phases (PI and PM) for various doping levels ($x=0,0.011,0.028,0.056$) measured at different temperatures (indicated by blue (dark gray) dots in Fig. \ref{fig:diagram}) and pressure (green (light gray) dots). 
In the framework of our previous theoretical understanding of V $K$ pre-edge XAS spectra\cite{Rodolakis2010}, we explore the changes in the electronic structure across the phase transition and its variations with external parameters (temperature, doping and pressure) within each phase. To complete the description of the phase diagram, X-ray diffraction (XRD) experiments under pressure and temperature have been also carried out.

The paper is organized as follows: In Sec.~\ref{sec:exp}, the experimental details of the XAS measurements are described. General features of the XAS spectra through the MIT are discussed in Sec.~\ref{sec:mit}. Spectral changes as a function of temperature (T), doping ($x$) and pressure (P), within a given phase, are presented and compared in Sec.~\ref{sec:xas}. A discussion is finally provided in Sec.~\ref{sec:discussion}.


\section{Experimental}
\label{sec:exp}

The experiments were performed on the inelastic x-ray scattering (IXS) beamlines ID26 and BM30 at the European Synchrotron Radiation Facility (ESRF). The X-ray energies were selected by a cryogenically-cooled double-crystal monochromator equipped with Si(311) (Si(220)) crystals at ID26 (BM30), which provided an energy bandwidth of 160~meV (280~meV) at the V K-edge. XRD measurements as a function of temperature and pressure have been performed on the CRISTAL beamline at SOLEIL.

To obtain an improved intrinsic resolution, the XAS spectra were acquired in the so-called partial fluorescence yield (PFY) mode \cite{Groot2001}. It consists of monitoring the intensity of the V-K$\alpha$ ($2p\rightarrow 1s$) emission line as the incident energy is swept across the absorption edge. With respect to standard XAS, the PFY mode provides higher resolution absorption spectra, as they are partly free from core-hole lifetime broadening effects. The gain is particularly appreciable at the K pre-edge of transition metals as the $1s$ core-hole is extremely short lived with respect to the $2p$ core-hole, and the dipolar tail strongly reduced. The V-K$\alpha$ line was measured with the help of an IXS Rowland-circle spectrometer equipped with a Ge(331) spherically bent analyzer with a bending radius of 860 cm. The same analyzer was used for all the PFY-XAS measurements.

We used powder specimens prepared from high quality (V$_{1-x}$Cr$_x$)$_2$O$_3$ single crystals from Purdue University, with various doping levels in the PM ($x=0$) and PI phases ($x=0.011,0.028,0.056$). The powder samples were compressed into pellets, well adapted to the fluorescence yield detection, occasionally with addition of boron nitride (BN) to reduce their density. A closed cycle cryocooler was installed on the IXS spectrometer to control the temperature. When applying an external pressure, the PFY mode turned out to be difficult to implement because of the weak fluorescence signal through the pressure cell, and we consequently  opted for standard XAS measurements in transmission mode. To maximize the throughput, the powders ($x=0.028$) were loaded in a diamond anvil cell equipped with composite anvils made of a perforated diamond capped with a 500-$\mu m$ thin diamond anvil; we used a low resolution Si(111) monochromator (instead of Si(311)) to improve the signal level. Pressure was measured in-situ by standard ruby fluorescence techniques. We used silicon oil as pressure transmitting medium in the cell. It should be noted that the fact that both K-edge XAS and XRD are based on the use of hard X-rays (and hence compatible with diamond pressure cells), which was essential in terms of varying not only temperature and doping but also pressure during our measurements. 
 
\begin{figure}[!tb]
\includegraphics[width=1\linewidth]{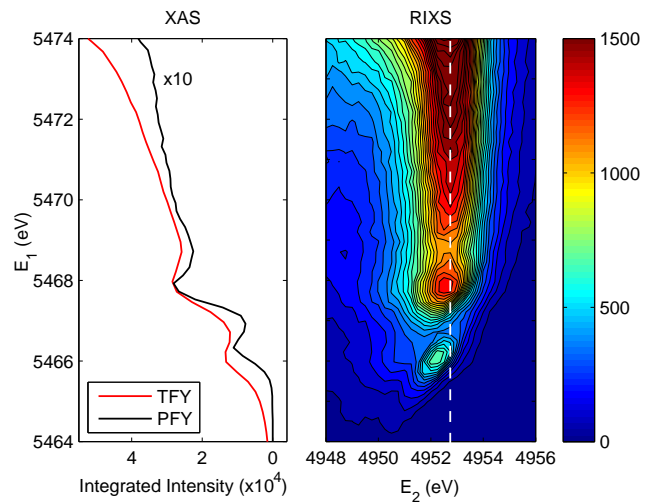}
\caption{(Color online) (a) V $K$ pre-edge X-ray absorption spectra in a \sesC\ single crystal measured at room conditions in the partial (PFY) and total (TFY) fluorescence modes; (b) RIXS surface as a function of incident (E$_{1}$) and emitted (E$_{2}$) photon energy in the vicinity of the pre-edge region.} 
\label{fig:RXES}
\end{figure}

We would like to point out that the PFY spectrum is nothing else but a specific cut through the general resonant inelastic X-ray scattering (RIXS) surface, where the spectral intensity is measured in resonant conditions as a function of both emitted and incident energies. Fig.~\ref{fig:RXES} shows a typical RIXS measurement in \sesC\ measured at ID26. The incident energy is here limited to the pre-edge region, while the emitted energy is monitored in the vicinity of the K$\alpha_1$ emission line. The left panel displays the XAS spectra retrieved from the RIXS intensity in both partial (PFY) and total (TFY) fluorescence yield mode. The latter was obtained by integrating the RIXS intensity over the entire emitted energy range, while the former is a cut over a 0.5 eV window around the K$\alpha_1$ emission line (white dash on the right panel). The sharpening effect is evident in the PFY acquisition mode. 
No missing  spectral weight is observed in the PFY spectrum of \sesC\ which is found to be almost identical, except for the enhanced resolution effects, to the standard X-ray absorption.
On the other hand, the choice of the PFY emission energy affects the positions of the lowest energy peak structure, as its maximum intensity in the RIXS surface is located slightly out of the K$\alpha_1$ emission line. Such artificial shift of 0.2 eV of the low-energy peak has been taken into account in the analysis, and it has therefore no impact on the final results of our study.

\begin{figure}[!tb]
\centering
\includegraphics[width=1\linewidth]{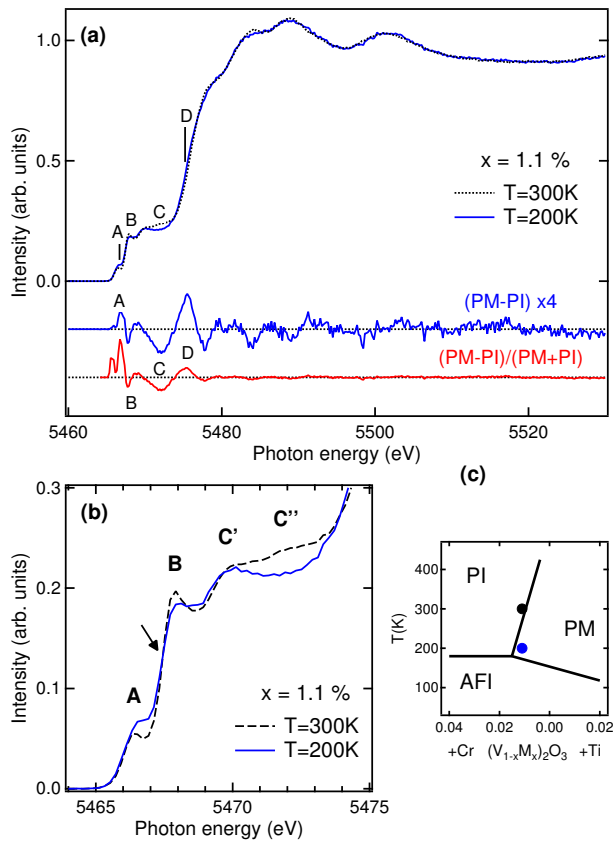}
\caption{(Color online) (a) V $K$-edge X-ray absorption spectra in a \sesC\ powder sample measured as a function of temperature in the PI (300~K, dashed line) and PM (200~K, solid line) phases in PFY mode; below, absolute and relative spectral differences (blue and red line, respectively); (b) pre-edge region; (c) corresponding points in the phase diagram.} 
\label{fig:mit}
\end{figure}

\section{Spectral features of the V K-edge}
\label{sec:mit}

Figure~\ref{fig:mit} shows the isotropic V K-edge PFY absorption spectra through the T-induced MIT. The full spectra are displayed in Fig.~\ref{fig:mit}(a) for both the PM (200 K) and PI (300 K) phase in a $x=0.011$ powder sample. The spectra were normalized at the inflection point of the first EXAFS oscillations. They can be decomposed in two spectral regions: the former is the main edge (above the onset (D) at 5475~eV), corresponding to $1s\rightarrow 4p$ dipolar transitions; the latter is the pre-edge (at lower energy), mainly corresponding to $1s\rightarrow 3d$ quadrupolar transitions, shown in Fig.~\ref{fig:mit}(b) on an expanded energy scale. 
Visible spectral changes are observed through the transition, as it is clearly shown by the spectral difference (PM-PI). The integrated intensity of the difference curve averages to zero over the whole spectra, but also in the pre-edge region (from 5460~eV to 5475~eV), suggesting a conservation of the spectral weight (SW) through the transition within the pre-edge.
A visual inspection of the relative spectral differences shows that these changes occur principally in the pre-edge region (peaks A,B, and area C) and at higher energies at the rising edge (D); beyond the main edge, the spectra are almost indistinguishable. As both paramagnetic phases have the same crystallographic structure, we can conclude that these differences are only due to modifications in the electronic structure occurring at the MIT under the effect of electronic correlations.

In this study, we will more specifically focus on the pre-edge region, and on how it evolves through the MIT as a function of the various external parameters. The pre-edge regions for both paramagnetic phases present well defined spectral features, labelled (A,B,C), which vary in intensity and shape as the system is driven through the MIT. The broad feature C can be further split into two sub structures at about 5470~eV (C') and 5472~eV (C''), the latter exhibiting the largest change with an inversion of curvature at the transition.

Understanding the spectral features of the whole pre-edge region in \sesX\ remains a complex task. 
First, the V sites lack inversion symmetry along the c-axis resulting in a slight on-site mixing of $3d$ and $4p$ vanadium states. As a consequence, dipole transitions to the formerly ``pure'' $3d$ states are possible to a certain extent (note that such mixing leads to interference effects that can be visible in elastic scattering experiments~\cite{Hansmann_unpublished}).
Second, the ground states of the PM and the PI phases is strongly multi-orbital in nature as it involves a mixing of the $e_g^\pi$ and $a_{1g}$ orbitals. The PI-PM transition is characterized by a change in the occupancy ratio between the ($e_{g}^{\pi}$,$a_{1g}$) and ($e_{g}^{\pi}$,$e_{g}^{\pi}$) states, which is markedly smaller in the PI with respect to the PM phase~\cite{Keller2004,Park2000}. The detailed analysis of the lowest energy pre-edge 
features of the V K-edge of \sesX\ has been the object of our previous paper~\cite{Rodolakis2010}. 
%

Starting from the higher energy region of the pre-edge, it may be tempting to associate peak C to non-local $1s\rightarrow3d$ dipole excitations to a neighbouring V atom~\cite{Gougoussis2009}. Such transitions are associated with the $4p$-$O$-$3d$ intersite hybridization, which is expected to sensitively depend on the metal-ligand distance~\cite{Vanko2008}. However, although the insulator to metal transition in the paramagnetic phase is isostructural, the lattice parameters change discontinuously at the transition. In particular, the basal $a$ axis decreases~\cite{McWhan1970}, resulting in a reduced $V-O$ bond. This should be related to the strong change observed in the C'' intensity, as we will further investigate in Sec.~\ref{sec:xaspression}: our experiments reveal indeed a relation between changes in the lattice parameters and C'' intensity.  Nevertheless, a simple interpretation of these feature based on non-local terms is not evident, since a reduced $V-O$ bond should imply a larger hybridization. This would be consequently expected to induce a larger intensity, which is definitely not the case in our spectra. 

An accurate theoretical description of the higher energy pre-edge region, corresponding to excitations involving the $e_g^\sigma$ states, is highly non-trivial. The reason for this is that the character of the $e_g^\sigma$ states is neither entirely itinerant nor entirely localised which, consequently, prohibits the use of established tools tailored either for one limit or the other.
The interpretation of peak A and B, which correspond to excitonic features involving the \tg-states, is instead much clearer and  certainly more informative about the the low-energy physics of the MIT in V$_2$O$_3$. Specifically, with the help of 
multiplet calculations in the configuration interaction (CI) scheme, combined with LDA+DMFT calculations~\cite{Rodolakis2010}, we confirmed the change in the (\eg,\eg):(\eg,\ag) occupancy ratio from $50:50$ in the PM phase to $35:65$ in PI phase, in good agreement with Ref.~\onlinecite{Park2000}.  
The isotropic CI-based calculated XAS spectra agree well with the experimental data as for both the energy splitting of the features A and B and the ratio of their SW which increases in the PM phase. 
As a result, peaks A and B in the pre-edge region have to be related to incoherent (local) excitations involving $t_{2g}$ degrees of freedom~\cite{Rodolakis2010}. 

An important outcome from these calculations is that the intensity ratio of the first two pre-edge features (i.e. the ratio between peaks A and B) is a key spectral parameter related to the differences between PM and PI: the larger the ratio between the spectral weight of A and B, the larger the $a_{1g}$ orbital occupation.\\

In our previous work~\cite{Rodolakis2010}, we focused our interest on the differences among P, T and $x$-induced transitions in \sesX, but interesting insight can be also gained by analyzing differences in XAS spectra as a function of these external parameters within a given phase. In this paper, we explore in detail how the XAS spectroscopic features in pre-edge (A,B,C' and C'') evolve while moving across the phase diagram,
and discuss how these changes are related to the physics of the system.


\section{XAS spectra across the phase diagram}
\label{sec:xas}

In this section, we present high resolution XAS measurements of \sesX \ over an extensive portion of its phase diagram. The effects of each thermodynamic parameter (temperature, doping, pressure) will be discussed in the forthcoming individual subsections. For sake of clarity, the main spectroscopic results will be then presented in a summary figure at the end of the paper  (Figure~\ref{fig:analysis}), to give a general picture of their evolution. In parallel, a summary figure for the lattice parameters will be also presented at the end (Figure~\ref{fig:lattice}), where the physical implications of our results will be discussed. We found this very useful, since the subtle interplay between electronic and lattice degrees of freedom plays a key role in the physics of \sesX\ : in the effort of providing the reader with as much information as possible, we collected in Figure~\ref{fig:lattice} the results of X-ray diffraction measurements from previous authors\cite{McWhan1970}, together with new data of ours.   
We recall here the main trends of the structural behavior of \sesX\, which are summerized in Figure~\ref{fig:lattice} for the PI-PM transition as a function of temperature, doping and pressure (left and middle panels). For all the thermodynamic parameters under consideration, at the PI-PM transition an increase can be observed for $c$, while $a$ decreases, resulting in a jump of the $c/a$ ratio. Data obtained by crossing the transition in temperature are remarkably identical to those measuring across the doping-induced transition from Ref~\onlinecite{McWhan1970}. On the other hand, measurements under pressure show the same trend, but with a smaller discontinuity.

These structural effects should be kept in mind during the analysis of the electronic properties, based on the XAS data presented hereafter. 

\subsection{Spectral changes with temperature and doping}
\label{sec:xastemperature}

\begin{figure}[!tb]
\centering
\includegraphics[width=1\linewidth]{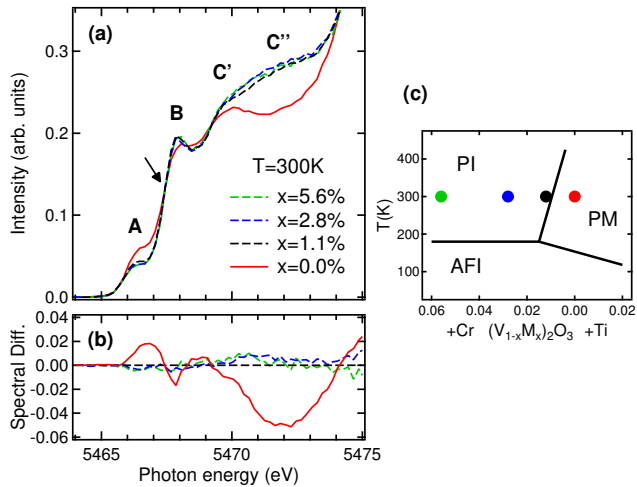}
\caption{(Color online) V-$K$ pre-edge X-ray absorption spectra for a \sesX \ powder sample at room conditions as a function of doping ($x=0,0.011,0.028,0.056$) in PFY mode; (b) spectral differences, with respect to the $x=0.011$ spectra; (c) corresponding points in the phase diagram.} 
\label{fig:dopage}
\end{figure}

Fig.~\ref{fig:dopage}(a) shows the spectra of \sesX\ for various doping levels ($x=0,0.011,0.028,0.056$) at ambient conditions. As shown in the phase diagram (Fig.~\ref{fig:dopage}(c)), the sample is stabilized in the PM phase for $x=0$ and in the PI phase for the other doping levels. To help visualize the spectral changes as a function of external parameters, we plotted in Fig.~\ref{fig:dopage}(b) the differences between the different data sets and the one for $x=0.011$.
We outlined in our previous paper~\cite{Rodolakis2010} the remarkable similarity between temperature and doping-induced changes through the MIT in the pre-edge XAS features: at low energy (A,B), the spectra measured through the doping induced MIT are identical within the experimental uncertainty to those measured through the T-driven transition, with a comparable A/B intensity ratio variation of $\approx15\%$.

Here we can see that the difference spectra in Fig.~\ref{fig:dopage}(b) shows barely any $x$-dependence in the PI phase, which is consistent with the recent optical conductivity measurements performed by Lupi \textit{et al.}\cite{Lupi2010}. 
In the frame of our interpretation, this indicates that the orbital occupation is rather insensitive to the doping level within the PI phase, in good agreement with the results obtained by Park \textit{et al.}~\cite{Park2000}. Consequently, local incoherent excitations probed by K and L$_{2,3}$ edge XAS on powder samples do not seem to be directly affected by local distortions induced by chemical substitution~\cite{Frenkel2006}, even though an anomalous behavior has recently been observed in the C',C'' spectral region for single crystals~\cite{Pease2011}, in parts of the phase diagram where, as recently evidenced, the PI and PM phases coexist~\cite{Lupi2010}.\\

\begin{figure}[!tb]
\centering
\includegraphics[width=0.95\linewidth]{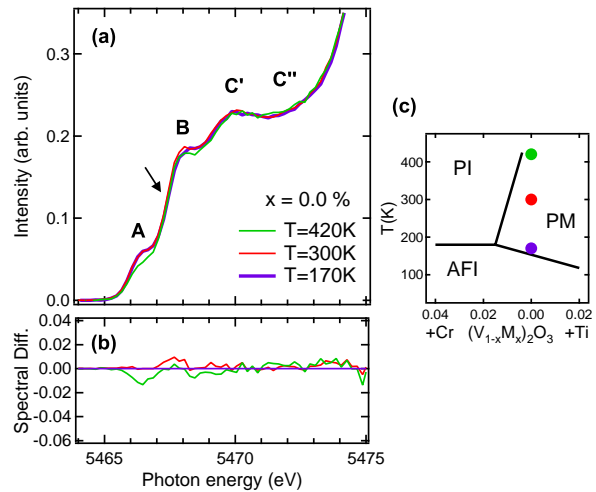}
\caption{(Color online) V-$K$ pre-edge X-ray absorption spectra for a \sesV\ (PM) powder sample as a function of temperature in PFY mode (T=170K,300K,420K); (b) spectral differences, with respect to the low temperature spectra; (c) corresponding points in the phase diagram.} 
\label{fig:temperature1}
\end{figure}

From this analysis, it appears that both A/B and C change at the PI/PM 
transition, while both remain constant within the Mott insulating (PI) phase. 
Their behavior within the correlated metallic (PM) phase is instead  
not as trivial - suggesting that the direct dependence  
on electron correlations is more pronounced for A/B than for C.  

Figure~\ref{fig:temperature1} displays the pre-edge region of XAS spectra in paramagnetic metallic \sesV\ as a function of temperature. Only negligible spectral changes are observed in the pre-edge structures between 300 K and 170 K; at higher temperature, 420 K, more pronounced modifications are visible for the low energy excitations (in particular for A). Thus, while the intensity ratio $I(A)/I(B)$ remains nearly constant when the sample is heated from 170K to 300K, a drop of $\approx10\%$ is observed between 300K and 420K, indicating a significative change in the orbital occupation. This change is comparable with the one measured across the (T,$x$)-MIT, as can be seen by comparing the evolution of the intensity ratio in Figure~\ref{fig:analysis} (e) and (f). It could be related to the proximity of the cross-over region which results in a variation of the conductivity in pure \sesV\ as a function of temperature~\cite{Kuwamoto1980}: starting from low temperature, the resistivity shows a monotonic decrease which is markedly enhanced while approaching the cross-over region around 425~K \cite{Baldassarre2008}. This is accompanied by a slight change in the lattice parameters, reproduced in Figure~\ref{fig:lattice} (right panel), which show as well an anomaly near to the cross-over region~\cite{McWhan1970}. Altogether, these results indicate that at the edge of the crossover region at 420K, differences between PI and PM already start to blur out.

A detailed XAS study at high temperature (above 420~K) on both sides of the transition line should certainly provide interesting information on the peculiar electronic properties in the crossover region, where a clear distinction between metal and insulator phases no longer exists~\cite{Mo2004}: in fact, as it can be 
seen in Figure~\ref{fig:lattice}, the values of c/a in the CO region tend 
progressively to the range corresponding to the PI phase.\\

Overall, except for these effects occurring when entering the crossover, no significant changes are visible in the low energy XAS spectral features when probing different (T,$x$) points in a given thermodynamic phase. 
By changing these external parameters, one would expect the coherent part of the electron density of states to be affected by the subtle effects linked to disorder. In fact, using bulk sensitive photoemission it has been shown that the quasiparticle peak observed in the paramagnetic metallic phase presents a clear T-dependence while the lower Hubbard band remains almost unchanged \cite{Rodolakis2009}. This behavior is also observed as a function of doping level \cite{Mo2006}: at room temperature, the coherent SW is strongly reduced as the doping level is increased, while the lower Hubbard band is virtually not affected. This confirms that even the low energy spectral features in the XAS V K pre-edge region are directly linked only to the incoherent part of the electron density of states.

\begin{figure}[!tb]
\centering
\includegraphics[width=0.95\linewidth]{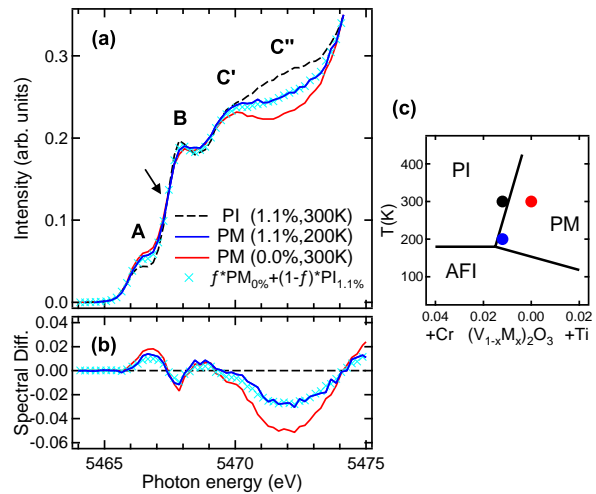}
\caption{(Color online) V-$K$ pre-edge X-ray absorption spectra for a \sesX\ powder sample as a function of temperature and doping level in PFY mode ($x=0,0.011$ and T=200K,300K); crosses represent a linear combination of the room temperature spectra (PM$_{0\%}$ and PI$_{1.1\%}$) for a value of $f=0.54$; (b) PI-PM spectral differences; (c) corresponding points in the phase diagram.} 
\label{fig:badmetal}
\end{figure}


Nevertheless, significant changes in the correlated metal XAS spectral yield appear in one part of the PM region of the phase diagram: 
as shown in the Figure~\ref{fig:badmetal}, a marked difference is visible between the C'' feature for $x=0.011$ at T=200 K, and $x=0$ at T=300 K. This difference can be linked to the bad-metallic behavior observed from resistivity \cite{Kuwamoto1980} and optical conductivity \cite{Lupi2010} measurements in \sesC, differently from \sesV. As demonstrated in Ref.~\onlinecite{Lupi2010}, this bad metallic behavior can be explained by the presence of phase separation, characterized by the presence of coexisting metallic and insulating domains. This was unambiguously and directly demonstrated using a spectromicroscopic approach~\cite{Grego99}, a method with a well established capability of resolving metallic from insulating microscopic regions~\cite{Gunther97}. 
Subsequently, the effects of this phase separation on the XAS spectra of 
\sesX\ single crystals were analyzed in details~\cite{Pease2011}, showing an anomalous behavior of a feature corresponding to the spectral region that we 
indicate with C'' and that has been related to the presence of highly distorted
VO$_6$ octahedra. 
 In our case, where only powder samples are considered, we notice that the doped metallic spectrum (PM$_{1.1\%}$) can be satisfactorily reproduced with a linear combination of the (PM$_{0\%}$) and (PI$_{1.1\%}$) spectra: \[f*PM_{0\%}+(1-f)*PI_{1.1\%}\]
The best fit is obtained for $f=0.54$ (crosses in Fig.~\ref{fig:badmetal}). This value is in good agreement with the volume fraction $f=0.42$ determined using a spatially resolved method like scanning photoelectron microscopy~\cite{Eich2000}, and successfully used for theoretical calculations of the spatially averaged optical conductivity~\cite{Lupi2010} (the larger value obtained here for $f$ can be explained considering the slightly lower temperature for the measurements presented here). 
We would also like to emphasize that these results have been consistently obtained with different sets of measurements performed on different synchrotron radiation beamlines, thanks also to a rigourous control of the experimental parameters (temperature, pressure, hysteresis at the phase transition, etc.).

\begin{figure}[!tb]
\centering
\includegraphics[width=0.95\linewidth]{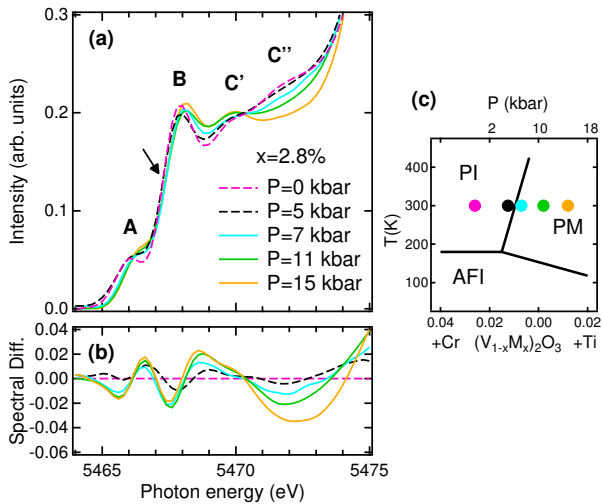}
\caption{(Color online) V-$K$ pre-edge X-ray absorption spectra in a \sesX \ ($x=0.028$) powder sample measured as a function of pressure in the PI (P $\leq$ 5 kbar, dashed lines) and PM (P $\geq$ 7 kbar, solid lines) phases; ; (b)  spectral differences, with respect to the room pressure spectra; the pressure scale in the phase diagram (c) refers to the $x=0.028$ doping.} 
\label{fig:pression}
\end{figure}

\subsection{Spectral changes with pressure}
\label{sec:xaspression}

Figure~\ref{fig:pression}(a) displays the pressure dependent pre-edge XAS spectra up to 15 kbar in a \sesX \ ($x=0.028$) powder sample; the MIT happens between 5 and 7~kbar (cf.\ Fig.~\ref{fig:lattice}). The spectra here obtained in transmission geometry have been deconvolved from the $1s$ Lorentzian lifetime broadening (1 eV FWHM) using the \emph{deconv} package from the GNXAS code~\cite{Filipponi1995,Filipponi1995a,Filipponi2000} so to match the improved resolution of the datasets measured by PFY-XAS. 

All the spectral features (A,B,C',C'') evolve under pressure in both paramagnetic phases, PI ($P\leq5$ kbar) and PM ($P\geq7$ kbar).
We observed a modification in the first two peaks intensity but these effects are not comparable to the abrupt modification of the spectral lineshape observed at the T or doping induced MIT: we measured the intensity ratio $I(A)/I(B)$ to raise of $15\%$ by crossing the transition with temperature or doping, versus only few per cent with pressure ($<2\%$). The largest changes are observed in the C spectral region. While C' slowly increases with pressure, C'' acquires a concave lineshape that progressively deepens with pressure as soon as the MIT is crossed. But the two regimes can be better visualized by inspecting the position of the inflection point between A and B as a function of pressure (cf. Fig~\ref{fig:analysis}(a)), which measures the rigid shift of the pre-edge with respect to the main edge. The discontinuous jump at 5 kbar clearly denotes the PI-PM transition line, while only negligible changes are visible within the two phases. 

Actually, the evolution of the P-dependent difference spectra (Fig~\ref{fig:pression}(b)) is rather continuous, and the P-induced MIT results in a rigid shift of the spectral weight of the first two peaks up in energy by $\sim$0.15~eV, as explained in our previous paper~\cite{Rodolakis2010}. These results, resumed in the Figure~\ref{fig:analysis}, are in striking contrast with what is seen as a function of (T,$x$).

\begin{figure}[!b]
\centering
\includegraphics[width=0.95\linewidth]{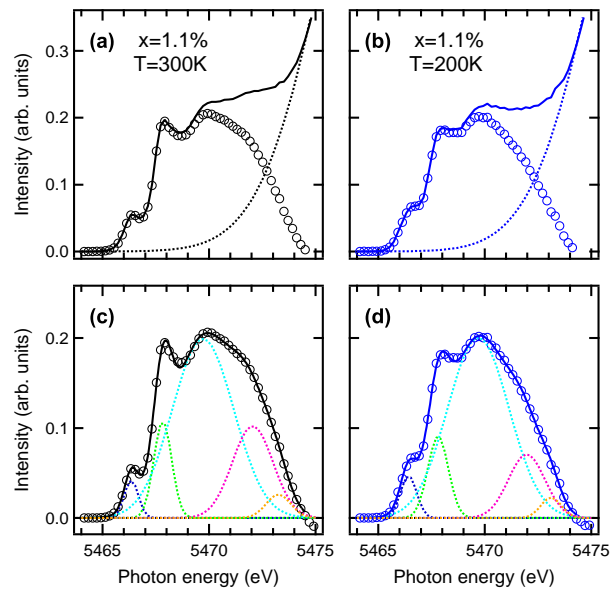}
\caption{(Color online) Illustration of the pre-edge fitting procedure for the data obtained on the \sesC\ sample across the temperature induced MIT: after subtraction of the dipolar tails (dotted lines, top panel), the remained features (circles) are fitted using five gaussian functions (dotted lines, bottom panel). The maxima of the first two gaussians corresponding to the peaks A and B are used to display the $I(A)/I(B)$ evolution in Figure~\ref{fig:analysis}(d-f).} 
\label{fig:fit}
\end{figure}

\section{Discussion}
\label{sec:discussion}

\begin{figure*}[!t]
\centering
\includegraphics[width=0.85\linewidth]{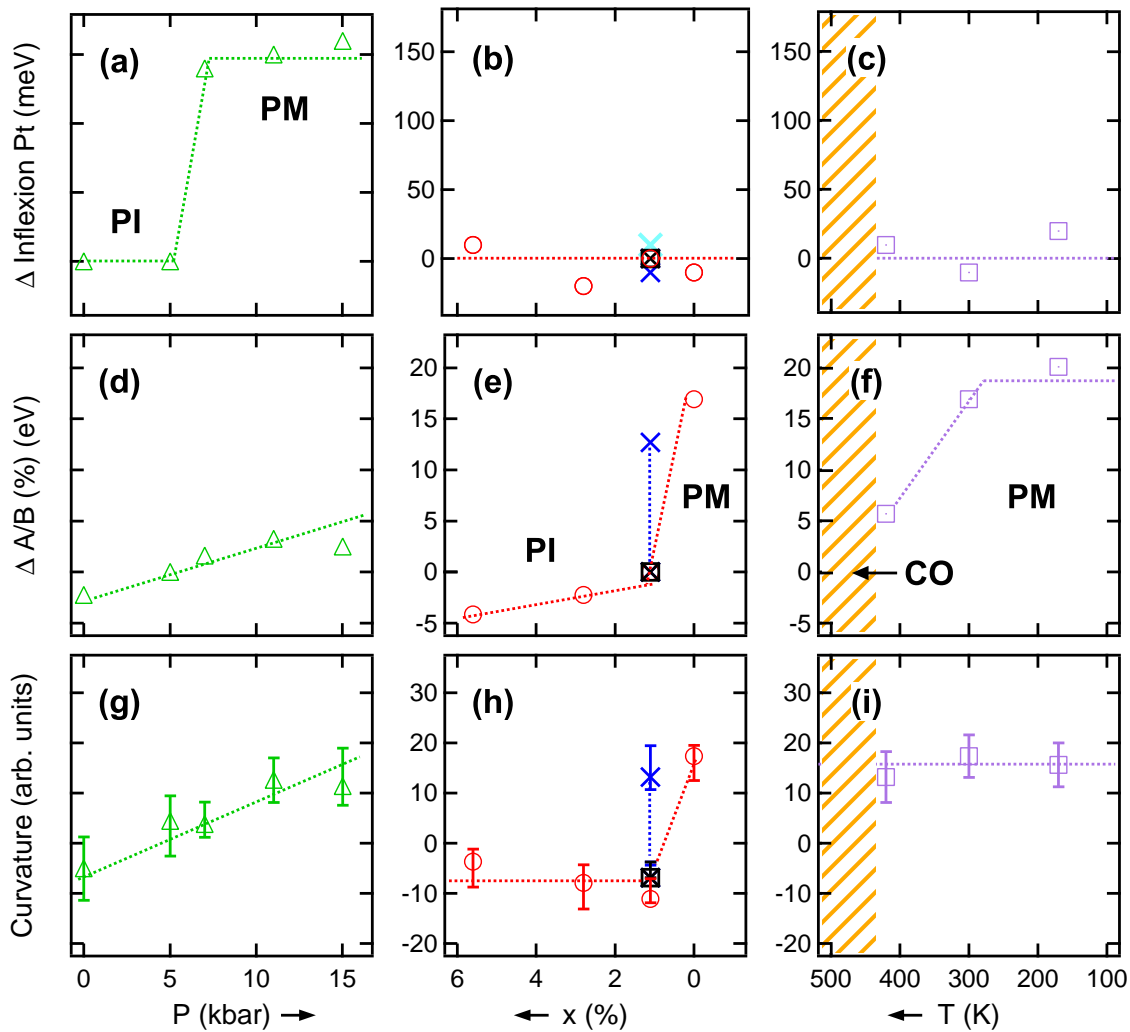}
\caption{(Color online) Pre-edge spectral feature evolution as a function of pressure ($x=2.8\%$, green triangles; the pressure scale refers to the $x=2.8\%$ doping), doping level (red circles) and temperature ($x=0\%$, purple square): (a-c) relative displacement of the inflection point between the first two peaks, indicated by the black arrow on the pre-edge spectra; (d-f) relative variation of the intensity ratio $I(A)/I(B)$; (g-i) inverse radius of curvature at the C'' point. Shaded areas indicate the crossover (CO) region. On the middle panel we present the data obtained in the \sesC\ sample across the temperature-induced MIT in the PI (300K, black crossed square) and PM phases (200K, dark blue cross); the light blue cross (panel (b)) corresponds to the data obtained in TFY mode, showing these results are independent of the detection mode.} 
\label{fig:analysis}
\end{figure*}

\begin{figure*}[!tb]
\centering
\includegraphics[width=0.95\linewidth]{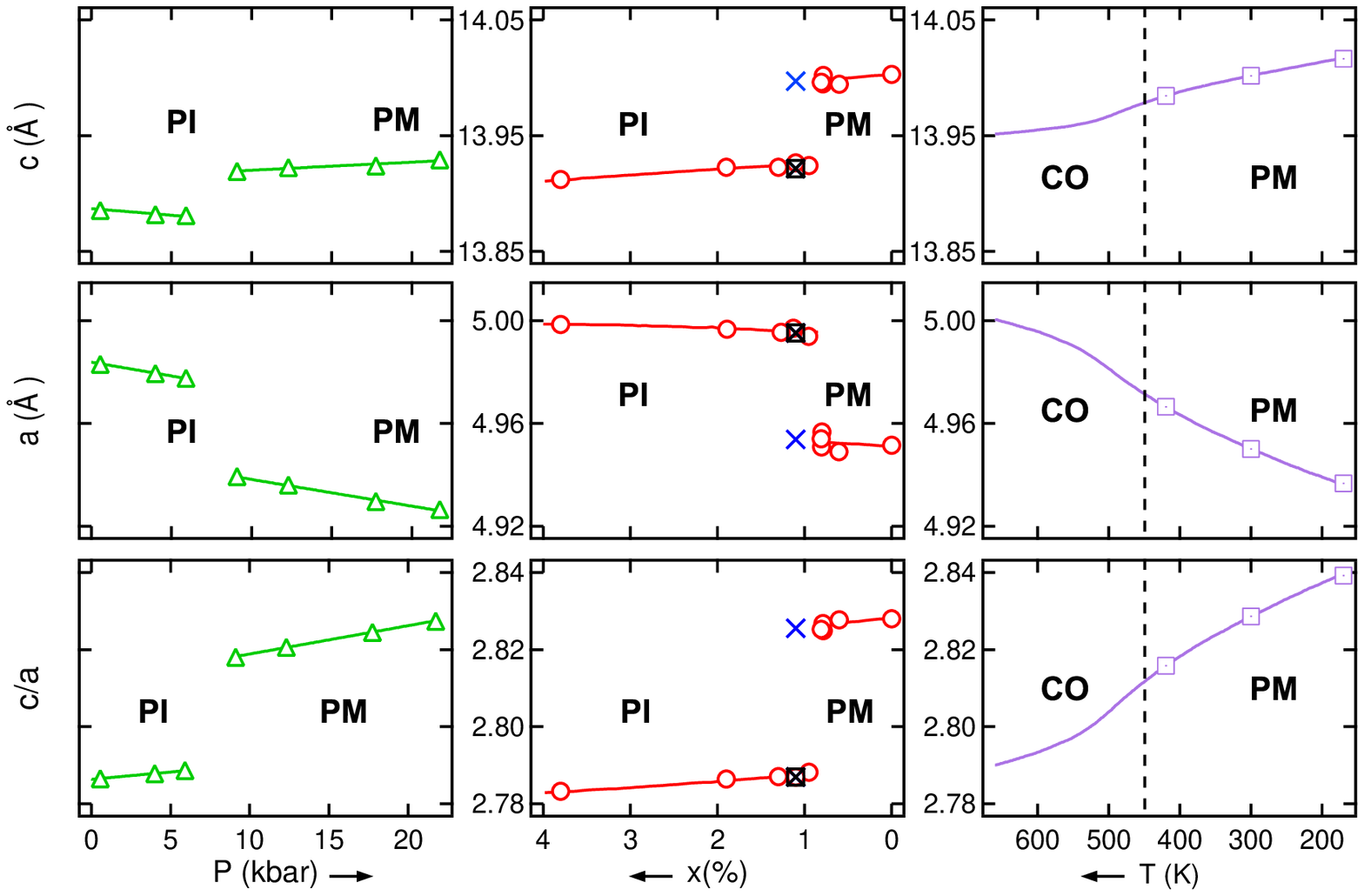}
\caption{(Color online) Lattice parameters as a function of pressure ($x=2.8\%$, green triangles; the pressure scale refers to the $x=2.8\%$ doping), doping level (red circles, from Ref.~\cite{McWhan1970}), and temperature ($x=0\%$, purple square, from Ref.~\cite{McWhan1970}). On the middle panel we also present the lattice parameters measured in a \sesC\ sample across the temperature-induced MIT in the PI (300K, black crossed square) and PM phases (200K, blue cross).} 
\label{fig:lattice}
\end{figure*}

To make the comparison between the different thermodynamic parameters easier, we have extracted the evolution of different spectral features. These results are summarized in Fig.~\ref{fig:analysis} which shows: (a-c) the relative displacement of the inflection point between the first two peaks (indicated by the black arrow on the pre-edge spectra figures); (d-f) the relative variation of the intensity ratio $I(A)/I(B)$, determined using the fitting procedure exposed in Fig.~\ref{fig:fit}; (g-i) the inverse radius of curvature at the C'' point. 

A comparative analysis of the effects of the various thermodynamic parameters leads to the following observations: \textit{i)} within a phase, the spectral features show barely any dependence as a function of temperature and doping level, in striking contrast with pressure; \textit{ii)} while (T,$x$)-induced transitions result in a discontinuous evolution of a specific spectral feature across the MIT, the pressure dependence of this spectral feature is virtually continuous, and vice versa. These results confirm the peculiar role of pressure in the phase diagram of \sesX.

As the pressure is increased, the continuous lineshape modifications of the pre-edge features for both paramagnetic phases (cf. Fig.~\ref{fig:analysis}(d,g)) can be first attributed to the variation of the lattice parameter under compression: applying an external pressure induces a variation of lattice parameters within both PM and PI, and not mainly at the transition as observed as a function the doping level (left vs. middle panel in Fig.~\ref{fig:lattice}). The resulting variation in the $U/W$ ratio changes the correlations effects, which one can observe in the pre-edge features.

This interpretation is no longer evident considering the strong evolution of the lattice parameters as a function of temperature in the metallic phase.
However, one should note that XAS is a very local probe, contrary to XRD. Understanding the exact interplay between electronic and lattice degrees of freedom would require a more local structural study, such as EXAFS. 

In Fig.~\ref{fig:analysis}(d) the $I(A)/I(B)$ ratio is displayed as a function of pressure. We observe a continuous increase of about $5\%$ over the entire pressure range. The absence of any discontinuity through the MIT suggests that this evolution is not related to the transition itself, as can be observed for temperature or doping, but rather to the pressure increase, regardless of the considered phase. In the frame of the arguments discussed above, this evidences a slight increase of the \ag\ occupation in the ground state while $U/W$ decreases by increasing P. This result is in agreement with the theoretical evolution predicted by Keller \textit{at al.} \cite{Keller2004}, and supports our statement that different mechanisms apply for the pressure and temperature (or doping) induced transitions~\cite{Rodolakis2010}.


\section{Conclusion}
\label{sec:conclusion}

We performed a high resolution x-ray absorption spectroscopy study at the V K-edge of \sesX\, to probe its electronic structure through the isostructural metal-insulator transition and across its phase diagram as a function of temperature, doping and pressure. 

In particular, our detailed analysis of the V K pre-edge region over an extensive set of data confirms our previous interpretation of the nature of the various spectral features: in particular, the ratio between A and B provides a direct indication of the orbital occupation of the $a_{1g}$ orbital, and is consequently directly related to electron correlation effects. For the interpretation of C', C'' further modeling efforts are necessary, since their character in between the itinerant and local limit has, so far, limited a profound theoretical description. It is however clear that the shape of this spectral region is extremely sensitive to the coexistence of the two phases~\cite{Pease2010} although the origin of this sensitivity (structural, screening effects) is not fully understood. 

By analyzing this spectroscopic information across the phase diagram, and by comparing it with novel and old XRD measurements, we were able to confirm with an extensive set of data the key role played by the interplay of electron correlations and slight structural changes in the physics of this model Mott-Hubbard system, since our hard X-ray based techniques allowed us to compare the effects of doping, temperature and pressure. This interplay gives support to some of the most recent theoretical models describing this system ~\cite{Keller2004,Poteryaev2007}, and explains two important effects of  general interest for the physics of the MIT of realistic multiband systems.

On one side, our comparative study among the different thermodynamic parameters confirms the singularity of pressure in the phase diagram of this prototype system. This is due to the fact that, compared to doping, the action of pressure accommodates the lattice mismatch between metal and insulator in a less discontinuous way: this translates into a smaller c/a jump at the transition, associated with an unchanged occupation but increased bandwidth of the $a_{1g}$ orbitals. This less evident discontinuity at the transition is associated to more continuous and evident changes in parts of the phase diagram which are far away from the transition.

On the other side, it is also clear that the tendency of the system towards separation and coexistence of phases is also related at the amount of lattice mismatch it has to accommodate at the transition. Comparing our results for the PM phase in two points of the phase diagram which are both close to the PM/PI phase transition, namely $x=0.011$, $T=200 K$ and $x=0$, $T=420 K$, it is clear that the former presents stronger evidence for phase coexistence: since the system has to accommodate a $c/a$ discontinuity of 0.04, in the proximity of the phase transition it will tend to form domains with the PI and PM lattice parameters ~\cite{Lupi2010} to accommodate this difference. For $x=0$, $T=420 K$ we are instead  very close to the critical point and the crossover region, and the structural differences between PM and PI are reduced, as it is evident from the corresponding $c/a$ values: the fact that there is less lattice mismatch to accommodate explains while in this point, close to a Mott instability, there is no evidence of the dramatic phase separation found for $x=0.011$, $T=200K$, even though the thermodynamic distance from the PM/PI phase transition is similar.  
Alternatively, and more fundamentally, we may regard the PI-PM mismatch as the strength of the ``first-order'' transition (where $c/a$ can be regarded as the order parameter), and we may observe that this strength reflects the tendency towards phase separation: at low temperatures, i.e., for $x=0.011$, $T=200K$, there is indeed such a strong tendency. 
In contrast at $x=0$, $T=420$ K, which is close to the critical
point at which the first-order transition line terminates, not only the lattice
discontinuity gets very small, but we also do not find any strong tendency
towards phase separation.

In conclusion, this study shows that it is important for strongly correlated materials to combine the study of structural and electronic properties, and that exploring the way the material evolves while approaching a phase transition, and not only while crossing it, is also a fundamental piece of information in the effort of understanding these systems. 

We acknowledge partial financial support from the RTRA "`Triangle de la Physique" and from a BQR of the Universit\'{e} Paris-Sud.  


\bibliography{biblio_v45}

\begin{thebibliography}{27}%
\makeatletter
\providecommand \@ifxundefined [1]{%
 \@ifx{#1\undefined}
}%
\providecommand \@ifnum [1]{%
 \ifnum #1\expandafter \@firstoftwo
 \else \expandafter \@secondoftwo
 \fi
}%
\providecommand \@ifx [1]{%
 \ifx #1\expandafter \@firstoftwo
 \else \expandafter \@secondoftwo
 \fi
}%
\providecommand \natexlab [1]{#1}%
\providecommand \enquote  [1]{``#1''}%
\providecommand \bibnamefont  [1]{#1}%
\providecommand \bibfnamefont [1]{#1}%
\providecommand \citenamefont [1]{#1}%
\providecommand \href@noop [0]{\@secondoftwo}%
\providecommand \href [0]{\begingroup \@sanitize@url \@href}%
\providecommand \@href[1]{\@@startlink{#1}\@@href}%
\providecommand \@@href[1]{\endgroup#1\@@endlink}%
\providecommand \@sanitize@url [0]{\catcode `\\12\catcode `\$12\catcode
  `\&12\catcode `\#12\catcode `\^12\catcode `\_12\catcode `\%12\relax}%
\providecommand \@@startlink[1]{}%
\providecommand \@@endlink[0]{}%
\providecommand \url  [0]{\begingroup\@sanitize@url \@url }%
\providecommand \@url [1]{\endgroup\@href {#1}{\urlprefix }}%
\providecommand \urlprefix  [0]{URL }%
\providecommand \Eprint [0]{\href }%
\providecommand \doibase [0]{http://dx.doi.org/}%
\providecommand \selectlanguage [0]{\@gobble}%
\providecommand \bibinfo  [0]{\@secondoftwo}%
\providecommand \bibfield  [0]{\@secondoftwo}%
\providecommand \translation [1]{[#1]}%
\providecommand \BibitemOpen [0]{}%
\providecommand \bibitemStop [0]{}%
\providecommand \bibitemNoStop [0]{.\EOS\space}%
\providecommand \EOS [0]{\spacefactor3000\relax}%
\providecommand \BibitemShut  [1]{\csname bibitem#1\endcsname}%
\let\auto@bib@innerbib\@empty
\bibitem [{\citenamefont {McWhan}\ and\ \citenamefont
  {Remeika}(1970)}]{McWhan1970}%
  \BibitemOpen
  \bibfield  {author} {\bibinfo {author} {\bibfnamefont {D.~B.}\ \bibnamefont
  {McWhan}}\ and\ \bibinfo {author} {\bibfnamefont {J.~P.}\ \bibnamefont
  {Remeika}},\ }\href@noop {} {\bibfield  {journal} {\bibinfo  {journal} {Phys.
  Rev. B}\ }\textbf {\bibinfo {volume} {2}},\ \bibinfo {pages} {3734} (\bibinfo
  {year} {1970})}\BibitemShut {NoStop}%
\bibitem [{\citenamefont {McWhan}\ \emph {et~al.}(1973)\citenamefont {McWhan},
  \citenamefont {Menth}, \citenamefont {Remeika}, \citenamefont {Brinkman},\
  and\ \citenamefont {Rice}}]{McWhan1973}%
  \BibitemOpen
  \bibfield  {author} {\bibinfo {author} {\bibfnamefont {D.~B.}\ \bibnamefont
  {McWhan}}, \bibinfo {author} {\bibfnamefont {A.}~\bibnamefont {Menth}},
  \bibinfo {author} {\bibfnamefont {J.~P.}\ \bibnamefont {Remeika}}, \bibinfo
  {author} {\bibfnamefont {W.~F.}\ \bibnamefont {Brinkman}}, \ and\ \bibinfo
  {author} {\bibfnamefont {T.~M.}\ \bibnamefont {Rice}},\ }\href@noop {}
  {\bibfield  {journal} {\bibinfo  {journal} {Phys. Rev. B}\ }\textbf {\bibinfo
  {volume} {7}},\ \bibinfo {pages} {1920} (\bibinfo {year} {1973})}\BibitemShut
  {NoStop}%
\bibitem [{\citenamefont {McWhan}\ \emph {et~al.}(1969)\citenamefont {McWhan},
  \citenamefont {Rice},\ and\ \citenamefont {Remeika}}]{McWhan1969}%
  \BibitemOpen
  \bibfield  {author} {\bibinfo {author} {\bibfnamefont {D.~B.}\ \bibnamefont
  {McWhan}}, \bibinfo {author} {\bibfnamefont {T.~M.}\ \bibnamefont {Rice}}, \
  and\ \bibinfo {author} {\bibfnamefont {J.~P.}\ \bibnamefont {Remeika}},\
  }\href@noop {} {\bibfield  {journal} {\bibinfo  {journal} {Phys. Rev. Lett.}\
  }\textbf {\bibinfo {volume} {23}},\ \bibinfo {pages} {1384} (\bibinfo {year}
  {1969})}\BibitemShut {NoStop}%
\bibitem [{\citenamefont {Kuwamoto}\ \emph {et~al.}(1980)\citenamefont
  {Kuwamoto}, \citenamefont {Honig},\ and\ \citenamefont
  {Appel}}]{Kuwamoto1980}%
  \BibitemOpen
  \bibfield  {author} {\bibinfo {author} {\bibfnamefont {H.}~\bibnamefont
  {Kuwamoto}}, \bibinfo {author} {\bibfnamefont {J.~M.}\ \bibnamefont {Honig}},
  \ and\ \bibinfo {author} {\bibfnamefont {J.}~\bibnamefont {Appel}},\
  }\href@noop {} {\bibfield  {journal} {\bibinfo  {journal} {Phys. Rev. B}\
  }\textbf {\bibinfo {volume} {22}},\ \bibinfo {pages} {2626} (\bibinfo {year}
  {1980})}\BibitemShut {NoStop}%
\bibitem [{\citenamefont {Ezhov}\ \emph {et~al.}(1999)\citenamefont {Ezhov},
  \citenamefont {Anisimov}, \citenamefont {Khomskii},\ and\ \citenamefont
  {Sawatzky}}]{Ezhov1999}%
  \BibitemOpen
  \bibfield  {author} {\bibinfo {author} {\bibfnamefont {S.~Y.}\ \bibnamefont
  {Ezhov}}, \bibinfo {author} {\bibfnamefont {V.~I.}\ \bibnamefont {Anisimov}},
  \bibinfo {author} {\bibfnamefont {D.~I.}\ \bibnamefont {Khomskii}}, \ and\
  \bibinfo {author} {\bibfnamefont {G.~A.}\ \bibnamefont {Sawatzky}},\
  }\href@noop {} {\bibfield  {journal} {\bibinfo  {journal} {Phys. Rev. Lett.}\
  }\textbf {\bibinfo {volume} {83}},\ \bibinfo {pages} {4136} (\bibinfo {year}
  {1999})}\BibitemShut {NoStop}%
\bibitem [{\citenamefont {Paolasini}\ \emph {et~al.}(1999)\citenamefont
  {Paolasini}, \citenamefont {Vettier}, \citenamefont {de~Bergevin},
  \citenamefont {Yakhou}, \citenamefont {Mannix}, \citenamefont {Stunault},
  \citenamefont {Neubeck}, \citenamefont {Altarelli}, \citenamefont {Fabrizio},
  \citenamefont {Metcalf},\ and\ \citenamefont {Honig}}]{Paolasini1999}%
  \BibitemOpen
  \bibfield  {author} {\bibinfo {author} {\bibfnamefont {L.}~\bibnamefont
  {Paolasini}}, \bibinfo {author} {\bibfnamefont {C.}~\bibnamefont {Vettier}},
  \bibinfo {author} {\bibfnamefont {F.}~\bibnamefont {de~Bergevin}}, \bibinfo
  {author} {\bibfnamefont {F.}~\bibnamefont {Yakhou}}, \bibinfo {author}
  {\bibfnamefont {D.}~\bibnamefont {Mannix}}, \bibinfo {author} {\bibfnamefont
  {A.}~\bibnamefont {Stunault}}, \bibinfo {author} {\bibfnamefont
  {W.}~\bibnamefont {Neubeck}}, \bibinfo {author} {\bibfnamefont
  {M.}~\bibnamefont {Altarelli}}, \bibinfo {author} {\bibfnamefont
  {M.}~\bibnamefont {Fabrizio}}, \bibinfo {author} {\bibfnamefont {P.~A.}\
  \bibnamefont {Metcalf}}, \ and\ \bibinfo {author} {\bibfnamefont {J.~M.}\
  \bibnamefont {Honig}},\ }\href@noop {} {\bibfield  {journal} {\bibinfo
  {journal} {Phys. Rev. Lett.}\ }\textbf {\bibinfo {volume} {82}},\ \bibinfo
  {pages} {4719} (\bibinfo {year} {1999})}\BibitemShut {NoStop}%
\bibitem [{\citenamefont {Park}\ \emph {et~al.}(2000)\citenamefont {Park},
  \citenamefont {Tjeng}, \citenamefont {Tanaka}, \citenamefont {Allen},
  \citenamefont {Chen}, \citenamefont {Metcalf}, \citenamefont {Honig},
  \citenamefont {de~Groot},\ and\ \citenamefont {Sawatzky}}]{Park2000}%
  \BibitemOpen
  \bibfield  {author} {\bibinfo {author} {\bibfnamefont {J.-H.}\ \bibnamefont
  {Park}}, \bibinfo {author} {\bibfnamefont {L.~H.}\ \bibnamefont {Tjeng}},
  \bibinfo {author} {\bibfnamefont {A.}~\bibnamefont {Tanaka}}, \bibinfo
  {author} {\bibfnamefont {J.~W.}\ \bibnamefont {Allen}}, \bibinfo {author}
  {\bibfnamefont {C.~T.}\ \bibnamefont {Chen}}, \bibinfo {author}
  {\bibfnamefont {P.}~\bibnamefont {Metcalf}}, \bibinfo {author} {\bibfnamefont
  {J.~M.}\ \bibnamefont {Honig}}, \bibinfo {author} {\bibfnamefont {F.~M.~F.}\
  \bibnamefont {de~Groot}}, \ and\ \bibinfo {author} {\bibfnamefont {G.~A.}\
  \bibnamefont {Sawatzky}},\ }\href@noop {} {\bibfield  {journal} {\bibinfo
  {journal} {Phys. Rev. B}\ }\textbf {\bibinfo {volume} {61}},\ \bibinfo
  {pages} {11506} (\bibinfo {year} {2000})}\BibitemShut {NoStop}%
\bibitem [{\citenamefont {Keller}\ \emph {et~al.}(2004)\citenamefont {Keller},
  \citenamefont {Held}, \citenamefont {Eyert}, \citenamefont {Vollhardt},\ and\
  \citenamefont {Anisimov}}]{Keller2004}%
  \BibitemOpen
  \bibfield  {author} {\bibinfo {author} {\bibfnamefont {G.}~\bibnamefont
  {Keller}}, \bibinfo {author} {\bibfnamefont {K.}~\bibnamefont {Held}},
  \bibinfo {author} {\bibfnamefont {V.}~\bibnamefont {Eyert}}, \bibinfo
  {author} {\bibfnamefont {D.}~\bibnamefont {Vollhardt}}, \ and\ \bibinfo
  {author} {\bibfnamefont {V.~I.}\ \bibnamefont {Anisimov}},\ }\href@noop {}
  {\bibfield  {journal} {\bibinfo  {journal} {Phys. Rev. B}\ }\textbf {\bibinfo
  {volume} {70}},\ \bibinfo {pages} {205116} (\bibinfo {year}
  {2004})}\BibitemShut {NoStop}%
\bibitem [{\citenamefont {Poteryaev}\ \emph {et~al.}(2007)\citenamefont
  {Poteryaev}, \citenamefont {Tomczak}, \citenamefont {Biermann}, \citenamefont
  {Georges}, \citenamefont {Lichtenstein}, \citenamefont {Rubtsov},
  \citenamefont {Saha-Dasgupta},\ and\ \citenamefont
  {Andersen}}]{Poteryaev2007}%
  \BibitemOpen
  \bibfield  {author} {\bibinfo {author} {\bibfnamefont {A.~I.}\ \bibnamefont
  {Poteryaev}}, \bibinfo {author} {\bibfnamefont {J.~M.}\ \bibnamefont
  {Tomczak}}, \bibinfo {author} {\bibfnamefont {S.}~\bibnamefont {Biermann}},
  \bibinfo {author} {\bibfnamefont {A.}~\bibnamefont {Georges}}, \bibinfo
  {author} {\bibfnamefont {A.~I.}\ \bibnamefont {Lichtenstein}}, \bibinfo
  {author} {\bibfnamefont {A.~N.}\ \bibnamefont {Rubtsov}}, \bibinfo {author}
  {\bibfnamefont {T.}~\bibnamefont {Saha-Dasgupta}}, \ and\ \bibinfo {author}
  {\bibfnamefont {O.~K.}\ \bibnamefont {Andersen}},\ }\href@noop {} {\bibfield
  {journal} {\bibinfo  {journal} {Phys. Rev. B}\ }\textbf {\bibinfo {volume}
  {76}},\ \bibinfo {eid} {085127} (\bibinfo {year} {2007})}\BibitemShut
  {NoStop}%
\bibitem [{\citenamefont {Rodolakis}\ \emph {et~al.}(2010)\citenamefont
  {Rodolakis}, \citenamefont {Hansmann}, \citenamefont {Rueff}, \citenamefont
  {Toschi}, \citenamefont {Haverkort}, \citenamefont {Sangiovanni},
  \citenamefont {Tanaka}, \citenamefont {Saha-Dasgupta}, \citenamefont
  {Andersen}, \citenamefont {Held}, \citenamefont {Sikora}, \citenamefont
  {Alliot}, \citenamefont {Iti{\'e}}, \citenamefont {Baudelet}, \citenamefont
  {Wzietek}, \citenamefont {Metcalf},\ and\ \citenamefont
  {Marsi}}]{Rodolakis2010}%
  \BibitemOpen
  \bibfield  {author} {\bibinfo {author} {\bibfnamefont {F.}~\bibnamefont
  {Rodolakis}}, \bibinfo {author} {\bibfnamefont {P.}~\bibnamefont {Hansmann}},
  \bibinfo {author} {\bibfnamefont {J.-P.}\ \bibnamefont {Rueff}}, \bibinfo
  {author} {\bibfnamefont {A.}~\bibnamefont {Toschi}}, \bibinfo {author}
  {\bibfnamefont {M.~W.}\ \bibnamefont {Haverkort}}, \bibinfo {author}
  {\bibfnamefont {G.}~\bibnamefont {Sangiovanni}}, \bibinfo {author}
  {\bibfnamefont {A.}~\bibnamefont {Tanaka}}, \bibinfo {author} {\bibfnamefont
  {T.}~\bibnamefont {Saha-Dasgupta}}, \bibinfo {author} {\bibfnamefont {O.~K.}\
  \bibnamefont {Andersen}}, \bibinfo {author} {\bibfnamefont {K.}~\bibnamefont
  {Held}}, \bibinfo {author} {\bibfnamefont {M.}~\bibnamefont {Sikora}},
  \bibinfo {author} {\bibfnamefont {I.}~\bibnamefont {Alliot}}, \bibinfo
  {author} {\bibfnamefont {J.-P.}\ \bibnamefont {Iti{\'e}}}, \bibinfo {author}
  {\bibfnamefont {F.}~\bibnamefont {Baudelet}}, \bibinfo {author}
  {\bibfnamefont {P.}~\bibnamefont {Wzietek}}, \bibinfo {author} {\bibfnamefont
  {P.}~\bibnamefont {Metcalf}}, \ and\ \bibinfo {author} {\bibfnamefont
  {M.}~\bibnamefont {Marsi}},\ }\href@noop {} {\bibfield  {journal} {\bibinfo
  {journal} {Phys. Rev. Lett.}\ }\textbf {\bibinfo {volume} {104}},\ \bibinfo
  {pages} {047401} (\bibinfo {year} {2010})}\BibitemShut {NoStop}%
\bibitem [{\citenamefont {de~Groot}(2001)}]{Groot2001}%
  \BibitemOpen
  \bibfield  {author} {\bibinfo {author} {\bibfnamefont {F.}~\bibnamefont
  {de~Groot}},\ }\href@noop {} {\bibfield  {journal} {\bibinfo  {journal}
  {Chemical Reviews}\ }\textbf {\bibinfo {volume} {101}},\ \bibinfo {pages}
  {1779} (\bibinfo {year} {2001})}\BibitemShut {NoStop}%
\bibitem [{\citenamefont {Hansmann}()}]{Hansmann_unpublished}%
  \BibitemOpen
  \bibfield  {author} {\bibinfo {author} {\bibfnamefont {P.}~\bibnamefont
  {Hansmann}},\ }\href@noop {} {\bibinfo  {journal} {unpublished}\
  }\BibitemShut {NoStop}%
\bibitem [{\citenamefont {Gougoussis}\ \emph {et~al.}(2009)\citenamefont
  {Gougoussis}, \citenamefont {Calandra}, \citenamefont {Seitsonen},
  \citenamefont {Brouder}, \citenamefont {Shukla},\ and\ \citenamefont
  {Mauri}}]{Gougoussis2009}%
  \BibitemOpen
\bibfield  {journal} {  }\bibfield  {author} {\bibinfo {author} {\bibfnamefont
  {C.}~\bibnamefont {Gougoussis}}, \bibinfo {author} {\bibfnamefont
  {M.}~\bibnamefont {Calandra}}, \bibinfo {author} {\bibfnamefont
  {A.}~\bibnamefont {Seitsonen}}, \bibinfo {author} {\bibfnamefont
  {C.}~\bibnamefont {Brouder}}, \bibinfo {author} {\bibfnamefont
  {A.}~\bibnamefont {Shukla}}, \ and\ \bibinfo {author} {\bibfnamefont
  {F.}~\bibnamefont {Mauri}},\ }\href@noop {} {\bibfield  {journal} {\bibinfo
  {journal} {Phys. Rev. B}\ }\textbf {\bibinfo {volume} {79}},\ \bibinfo
  {pages} {045118} (\bibinfo {year} {2009})}\BibitemShut {NoStop}%
\bibitem [{\citenamefont {Vanko}\ \emph {et~al.}(2008)\citenamefont {Vanko},
  \citenamefont {de~Groot}, \citenamefont {Huotari}, \citenamefont {Cava},
  \citenamefont {Lorenz},\ and\ \citenamefont {Reuther}}]{Vanko2008}%
  \BibitemOpen
  \bibfield  {author} {\bibinfo {author} {\bibfnamefont {G.}~\bibnamefont
  {Vanko}}, \bibinfo {author} {\bibfnamefont {F.~M.~F.}\ \bibnamefont
  {de~Groot}}, \bibinfo {author} {\bibfnamefont {S.}~\bibnamefont {Huotari}},
  \bibinfo {author} {\bibfnamefont {R.~J.}\ \bibnamefont {Cava}}, \bibinfo
  {author} {\bibfnamefont {T.}~\bibnamefont {Lorenz}}, \ and\ \bibinfo {author}
  {\bibfnamefont {M.}~\bibnamefont {Reuther}},\ }\href@noop {} {\bibfield
  {journal} {\bibinfo  {journal} {Cond. Mat.}\ ,\ \bibinfo {pages}
  {arvix:0802.2744}} (\bibinfo {year} {2008})}\BibitemShut {NoStop}%
\bibitem [{\citenamefont {Lupi}\ \emph {et~al.}(2010)\citenamefont {Lupi},
  \citenamefont {Baldassarre}, \citenamefont {Mansart}, \citenamefont
  {Perucchi}, \citenamefont {Barinov}, \citenamefont {Dudin}, \citenamefont
  {Papalazarou}, \citenamefont {Rodolakis}, \citenamefont {Rueff},
  \citenamefont {Iti\'{e}}, \citenamefont {Ravy}, \citenamefont {Nicoletti},
  \citenamefont {Postorino}, \citenamefont {Hansmann}, \citenamefont {Parragh},
  \citenamefont {Toschi}, \citenamefont {Saha-Dasgupta}, \citenamefont
  {Andersen}, \citenamefont {Sangiovanni}, \citenamefont {Held},\ and\
  \citenamefont {Marsi}}]{Lupi2010}%
  \BibitemOpen
  \bibfield  {author} {\bibinfo {author} {\bibfnamefont {S.}~\bibnamefont
  {Lupi}}, \bibinfo {author} {\bibfnamefont {L.}~\bibnamefont {Baldassarre}},
  \bibinfo {author} {\bibfnamefont {B.}~\bibnamefont {Mansart}}, \bibinfo
  {author} {\bibfnamefont {A.}~\bibnamefont {Perucchi}}, \bibinfo {author}
  {\bibfnamefont {A.}~\bibnamefont {Barinov}}, \bibinfo {author} {\bibfnamefont
  {P.}~\bibnamefont {Dudin}}, \bibinfo {author} {\bibfnamefont
  {E.}~\bibnamefont {Papalazarou}}, \bibinfo {author} {\bibfnamefont
  {F.}~\bibnamefont {Rodolakis}}, \bibinfo {author} {\bibfnamefont {J.-P.}\
  \bibnamefont {Rueff}}, \bibinfo {author} {\bibfnamefont {J.-P.}\ \bibnamefont
  {Iti\'{e}}}, \bibinfo {author} {\bibfnamefont {S.}~\bibnamefont {Ravy}},
  \bibinfo {author} {\bibfnamefont {D.}~\bibnamefont {Nicoletti}}, \bibinfo
  {author} {\bibfnamefont {P.}~\bibnamefont {Postorino}}, \bibinfo {author}
  {\bibfnamefont {P.}~\bibnamefont {Hansmann}}, \bibinfo {author}
  {\bibfnamefont {N.}~\bibnamefont {Parragh}}, \bibinfo {author} {\bibfnamefont
  {A.}~\bibnamefont {Toschi}}, \bibinfo {author} {\bibfnamefont
  {T.}~\bibnamefont {Saha-Dasgupta}}, \bibinfo {author} {\bibfnamefont {O.~K.}\
  \bibnamefont {Andersen}}, \bibinfo {author} {\bibfnamefont {G.}~\bibnamefont
  {Sangiovanni}}, \bibinfo {author} {\bibfnamefont {K.}~\bibnamefont {Held}}, \
  and\ \bibinfo {author} {\bibfnamefont {M.}~\bibnamefont {Marsi}},\
  }\href@noop {} {\bibfield  {journal} {\bibinfo  {journal} {Nat. Comm.}\
  }\textbf {\bibinfo {volume} {1}},\ \bibinfo {pages} {105} (\bibinfo {year}
  {2010})}\BibitemShut {NoStop}%
\bibitem [{\citenamefont {Frenkel}\ \emph {et~al.}(2006)\citenamefont
  {Frenkel}, \citenamefont {Pease}, \citenamefont {Budnick}, \citenamefont
  {Metcalf}, \citenamefont {Stern}, \citenamefont {Shanthakumar},\ and\
  \citenamefont {Huang}}]{Frenkel2006}%
  \BibitemOpen
  \bibfield  {author} {\bibinfo {author} {\bibfnamefont {A.~I.}\ \bibnamefont
  {Frenkel}}, \bibinfo {author} {\bibfnamefont {D.~M.}\ \bibnamefont {Pease}},
  \bibinfo {author} {\bibfnamefont {J.~I.}\ \bibnamefont {Budnick}}, \bibinfo
  {author} {\bibfnamefont {P.}~\bibnamefont {Metcalf}}, \bibinfo {author}
  {\bibfnamefont {E.~A.}\ \bibnamefont {Stern}}, \bibinfo {author}
  {\bibfnamefont {P.}~\bibnamefont {Shanthakumar}}, \ and\ \bibinfo {author}
  {\bibfnamefont {T.}~\bibnamefont {Huang}},\ }\href@noop {} {\bibfield
  {journal} {\bibinfo  {journal} {Phys. Rev. Lett.}\ }\textbf {\bibinfo
  {volume} {97}},\ \bibinfo {eid} {195502} (\bibinfo {year}
  {2006})}\BibitemShut {NoStop}%
\bibitem [{\citenamefont {Pease}\ \emph {et~al.}(2011)\citenamefont {Pease},
  \citenamefont {Frenkel}, \citenamefont {Krayzman}, \citenamefont {Huang},
  \citenamefont {Shanthakumar}, \citenamefont {Budnick}, \citenamefont
  {Metcalf}, \citenamefont {Chudnovsky},\ and\ \citenamefont
  {Stern}}]{Pease2011}%
  \BibitemOpen
  \bibfield  {author} {\bibinfo {author} {\bibfnamefont {D.}~\bibnamefont
  {Pease}}, \bibinfo {author} {\bibfnamefont {A.}~\bibnamefont {Frenkel}},
  \bibinfo {author} {\bibfnamefont {V.}~\bibnamefont {Krayzman}}, \bibinfo
  {author} {\bibfnamefont {T.}~\bibnamefont {Huang}}, \bibinfo {author}
  {\bibfnamefont {P.}~\bibnamefont {Shanthakumar}}, \bibinfo {author}
  {\bibfnamefont {J.}~\bibnamefont {Budnick}}, \bibinfo {author} {\bibfnamefont
  {P.}~\bibnamefont {Metcalf}}, \bibinfo {author} {\bibfnamefont
  {F.}~\bibnamefont {Chudnovsky}}, \ and\ \bibinfo {author} {\bibfnamefont
  {E.}~\bibnamefont {Stern}},\ }\href@noop {} {\bibfield  {journal} {\bibinfo
  {journal} {Phys. Rev. B}\ }\textbf {\bibinfo {volume} {83}},\ \bibinfo
  {pages} {085105} (\bibinfo {year} {2011})}\BibitemShut {NoStop}%
\bibitem [{\citenamefont {Baldassarre}\ \emph {et~al.}(2008)\citenamefont
  {Baldassarre}, \citenamefont {Perucchi}, \citenamefont {Nicoletti},
  \citenamefont {Toschi}, \citenamefont {Sangiovanni}, \citenamefont {Held},
  \citenamefont {Capone}, \citenamefont {Ortolani}, \citenamefont {Malavasi},
  \citenamefont {Marsi}, \citenamefont {Metcalf}, \citenamefont {Postorino},\
  and\ \citenamefont {Lupi}}]{Baldassarre2008}%
  \BibitemOpen
  \bibfield  {author} {\bibinfo {author} {\bibfnamefont {L.}~\bibnamefont
  {Baldassarre}}, \bibinfo {author} {\bibfnamefont {A.}~\bibnamefont
  {Perucchi}}, \bibinfo {author} {\bibfnamefont {D.}~\bibnamefont {Nicoletti}},
  \bibinfo {author} {\bibfnamefont {A.}~\bibnamefont {Toschi}}, \bibinfo
  {author} {\bibfnamefont {G.}~\bibnamefont {Sangiovanni}}, \bibinfo {author}
  {\bibfnamefont {K.}~\bibnamefont {Held}}, \bibinfo {author} {\bibfnamefont
  {M.}~\bibnamefont {Capone}}, \bibinfo {author} {\bibfnamefont
  {M.}~\bibnamefont {Ortolani}}, \bibinfo {author} {\bibfnamefont
  {L.}~\bibnamefont {Malavasi}}, \bibinfo {author} {\bibfnamefont
  {M.}~\bibnamefont {Marsi}}, \bibinfo {author} {\bibfnamefont
  {P.}~\bibnamefont {Metcalf}}, \bibinfo {author} {\bibfnamefont
  {P.}~\bibnamefont {Postorino}}, \ and\ \bibinfo {author} {\bibfnamefont
  {S.}~\bibnamefont {Lupi}},\ }\href@noop {} {\bibfield  {journal} {\bibinfo
  {journal} {Phys. Rev. B}\ }\textbf {\bibinfo {volume} {77}},\ \bibinfo {eid}
  {113107} (\bibinfo {year} {2008})}\BibitemShut {NoStop}%
\bibitem [{\citenamefont {Mo}\ \emph {et~al.}(2004)\citenamefont {Mo},
  \citenamefont {Kim}, \citenamefont {Allen}, \citenamefont {Gweon},
  \citenamefont {Denlinger}, \citenamefont {Park}, \citenamefont {Sekiyama},
  \citenamefont {Yamasaki}, \citenamefont {Suga}, \citenamefont {Metcalf},\
  and\ \citenamefont {Held}}]{Mo2004}%
  \BibitemOpen
  \bibfield  {author} {\bibinfo {author} {\bibfnamefont {S.-K.}\ \bibnamefont
  {Mo}}, \bibinfo {author} {\bibfnamefont {H.-D.}\ \bibnamefont {Kim}},
  \bibinfo {author} {\bibfnamefont {J.~W.}\ \bibnamefont {Allen}}, \bibinfo
  {author} {\bibfnamefont {G.-H.}\ \bibnamefont {Gweon}}, \bibinfo {author}
  {\bibfnamefont {J.~D.}\ \bibnamefont {Denlinger}}, \bibinfo {author}
  {\bibfnamefont {J.-H.}\ \bibnamefont {Park}}, \bibinfo {author}
  {\bibfnamefont {A.}~\bibnamefont {Sekiyama}}, \bibinfo {author}
  {\bibfnamefont {A.}~\bibnamefont {Yamasaki}}, \bibinfo {author}
  {\bibfnamefont {S.}~\bibnamefont {Suga}}, \bibinfo {author} {\bibfnamefont
  {P.}~\bibnamefont {Metcalf}}, \ and\ \bibinfo {author} {\bibfnamefont
  {K.}~\bibnamefont {Held}},\ }\href@noop {} {\bibfield  {journal} {\bibinfo
  {journal} {Phys. Rev. Lett.}\ }\textbf {\bibinfo {volume} {93}},\ \bibinfo
  {pages} {076404} (\bibinfo {year} {2004})}\BibitemShut {NoStop}%
\bibitem [{\citenamefont {Rodolakis}\ \emph {et~al.}(2009)\citenamefont
  {Rodolakis}, \citenamefont {Mansart}, \citenamefont {Papalazarou},
  \citenamefont {Gorovikov}, \citenamefont {Vilmercati}, \citenamefont
  {Petaccia}, \citenamefont {Goldoni}, \citenamefont {Rueff}, \citenamefont
  {Lupi}, \citenamefont {Metcalf},\ and\ \citenamefont
  {Marsi}}]{Rodolakis2009}%
  \BibitemOpen
  \bibfield  {author} {\bibinfo {author} {\bibfnamefont {F.}~\bibnamefont
  {Rodolakis}}, \bibinfo {author} {\bibfnamefont {B.}~\bibnamefont {Mansart}},
  \bibinfo {author} {\bibfnamefont {E.}~\bibnamefont {Papalazarou}}, \bibinfo
  {author} {\bibfnamefont {S.}~\bibnamefont {Gorovikov}}, \bibinfo {author}
  {\bibfnamefont {P.}~\bibnamefont {Vilmercati}}, \bibinfo {author}
  {\bibfnamefont {L.}~\bibnamefont {Petaccia}}, \bibinfo {author}
  {\bibfnamefont {A.}~\bibnamefont {Goldoni}}, \bibinfo {author} {\bibfnamefont
  {J.~P.}\ \bibnamefont {Rueff}}, \bibinfo {author} {\bibfnamefont
  {S.}~\bibnamefont {Lupi}}, \bibinfo {author} {\bibfnamefont {P.}~\bibnamefont
  {Metcalf}}, \ and\ \bibinfo {author} {\bibfnamefont {M.}~\bibnamefont
  {Marsi}},\ }\href@noop {} {\bibfield  {journal} {\bibinfo  {journal} {Phys.
  Rev. Lett.}\ }\textbf {\bibinfo {volume} {102}},\ \bibinfo {pages} {066805}
  (\bibinfo {year} {2009})}\BibitemShut {NoStop}%
\bibitem [{\citenamefont {Mo}\ \emph {et~al.}(2006)\citenamefont {Mo},
  \citenamefont {Kim}, \citenamefont {Denlinger}, \citenamefont {Allen},
  \citenamefont {Park}, \citenamefont {Sekiyama}, \citenamefont {Yamasaki},
  \citenamefont {Suga}, \citenamefont {Saitoh}, \citenamefont {Muro},\ and\
  \citenamefont {Metcalf}}]{Mo2006}%
  \BibitemOpen
  \bibfield  {author} {\bibinfo {author} {\bibfnamefont {S.-K.}\ \bibnamefont
  {Mo}}, \bibinfo {author} {\bibfnamefont {H.-D.}\ \bibnamefont {Kim}},
  \bibinfo {author} {\bibfnamefont {J.~D.}\ \bibnamefont {Denlinger}}, \bibinfo
  {author} {\bibfnamefont {J.~W.}\ \bibnamefont {Allen}}, \bibinfo {author}
  {\bibfnamefont {J.-H.}\ \bibnamefont {Park}}, \bibinfo {author}
  {\bibfnamefont {A.}~\bibnamefont {Sekiyama}}, \bibinfo {author}
  {\bibfnamefont {A.}~\bibnamefont {Yamasaki}}, \bibinfo {author}
  {\bibfnamefont {S.}~\bibnamefont {Suga}}, \bibinfo {author} {\bibfnamefont
  {Y.}~\bibnamefont {Saitoh}}, \bibinfo {author} {\bibfnamefont
  {T.}~\bibnamefont {Muro}}, \ and\ \bibinfo {author} {\bibfnamefont
  {P.}~\bibnamefont {Metcalf}},\ }\href@noop {} {\bibfield  {journal} {\bibinfo
   {journal} {Phys. Rev. B}\ }\textbf {\bibinfo {volume} {74}},\ \bibinfo {eid}
  {165101} (\bibinfo {year} {2006})}\BibitemShut {NoStop}%
\bibitem [{\citenamefont {Gregoratti}\ \emph {et~al.}(1999)\citenamefont
  {Gregoratti}, \citenamefont {Gunther}, \citenamefont {Kovac}, \citenamefont
  {Marsi}, \citenamefont {Phaneuf},\ and\ \citenamefont {Kiskinova}}]{Grego99}%
  \BibitemOpen
  \bibfield  {author} {\bibinfo {author} {\bibfnamefont {L.}~\bibnamefont
  {Gregoratti}}, \bibinfo {author} {\bibfnamefont {S.}~\bibnamefont {Gunther}},
  \bibinfo {author} {\bibfnamefont {J.}~\bibnamefont {Kovac}}, \bibinfo
  {author} {\bibfnamefont {M.}~\bibnamefont {Marsi}}, \bibinfo {author}
  {\bibfnamefont {R.}~\bibnamefont {Phaneuf}}, \ and\ \bibinfo {author}
  {\bibfnamefont {M.}~\bibnamefont {Kiskinova}},\ }\href@noop {} {\bibfield
  {journal} {\bibinfo  {journal} {Phys. Rev. B}\ }\textbf {\bibinfo {volume}
  {59}},\ \bibinfo {pages} {2018} (\bibinfo {year} {1999})}\BibitemShut
  {NoStop}%
\bibitem [{\citenamefont {Gunther}\ \emph {et~al.}(1997)\citenamefont
  {Gunther}, \citenamefont {Kolmakov}, \citenamefont {Kovac}, \citenamefont
  {Marsi},\ and\ \citenamefont {Kiskinova}}]{Gunther97}%
  \BibitemOpen
  \bibfield  {author} {\bibinfo {author} {\bibfnamefont {S.}~\bibnamefont
  {Gunther}}, \bibinfo {author} {\bibfnamefont {A.}~\bibnamefont {Kolmakov}},
  \bibinfo {author} {\bibfnamefont {J.}~\bibnamefont {Kovac}}, \bibinfo
  {author} {\bibfnamefont {M.}~\bibnamefont {Marsi}}, \ and\ \bibinfo {author}
  {\bibfnamefont {M.}~\bibnamefont {Kiskinova}},\ }\href@noop {} {\bibfield
  {journal} {\bibinfo  {journal} {Phys. Rev. B}\ }\textbf {\bibinfo {volume}
  {56}},\ \bibinfo {pages} {5003} (\bibinfo {year} {1997})}\BibitemShut
  {NoStop}%
\bibitem [{\citenamefont {Eich}\ \emph {et~al.}(2000)\citenamefont {Eich},
  \citenamefont {Herber}, \citenamefont {Groh}, \citenamefont {Stahl},
  \citenamefont {Heske}, \citenamefont {Marsi}, \citenamefont {Kiskinova},
  \citenamefont {Riedl}, \citenamefont {Fink},\ and\ \citenamefont
  {Umbach}}]{Eich2000}%
  \BibitemOpen
  \bibfield  {author} {\bibinfo {author} {\bibfnamefont {D.}~\bibnamefont
  {Eich}}, \bibinfo {author} {\bibfnamefont {U.}~\bibnamefont {Herber}},
  \bibinfo {author} {\bibfnamefont {U.}~\bibnamefont {Groh}}, \bibinfo {author}
  {\bibfnamefont {U.}~\bibnamefont {Stahl}}, \bibinfo {author} {\bibfnamefont
  {C.}~\bibnamefont {Heske}}, \bibinfo {author} {\bibfnamefont
  {M.}~\bibnamefont {Marsi}}, \bibinfo {author} {\bibfnamefont
  {M.}~\bibnamefont {Kiskinova}}, \bibinfo {author} {\bibfnamefont
  {W.}~\bibnamefont {Riedl}}, \bibinfo {author} {\bibfnamefont
  {R.}~\bibnamefont {Fink}}, \ and\ \bibinfo {author} {\bibfnamefont
  {E.}~\bibnamefont {Umbach}},\ }\href@noop {} {\bibfield  {journal} {\bibinfo
  {journal} {Thin Solid Films}\ }\textbf {\bibinfo {volume} {361-362}},\
  \bibinfo {pages} {258} (\bibinfo {year} {2000})}\BibitemShut {NoStop}%
\bibitem [{\citenamefont {Filipponi}\ and\ \citenamefont
  {Di~Cicco}(1995)}]{Filipponi1995}%
  \BibitemOpen
  \bibfield  {author} {\bibinfo {author} {\bibfnamefont {A.}~\bibnamefont
  {Filipponi}}\ and\ \bibinfo {author} {\bibfnamefont {A.}~\bibnamefont
  {Di~Cicco}},\ }\href@noop {} {\bibfield  {journal} {\bibinfo  {journal}
  {Phys. Rev. B}\ }\textbf {\bibinfo {volume} {52}},\ \bibinfo {pages} {15135}
  (\bibinfo {year} {1995})}\BibitemShut {NoStop}%
\bibitem [{\citenamefont {Filipponi}\ \emph {et~al.}(1995)\citenamefont
  {Filipponi}, \citenamefont {Di~Cicco},\ and\ \citenamefont
  {Natoli}}]{Filipponi1995a}%
  \BibitemOpen
  \bibfield  {author} {\bibinfo {author} {\bibfnamefont {A.}~\bibnamefont
  {Filipponi}}, \bibinfo {author} {\bibfnamefont {A.}~\bibnamefont {Di~Cicco}},
  \ and\ \bibinfo {author} {\bibfnamefont {C.~R.}\ \bibnamefont {Natoli}},\
  }\href@noop {} {\bibfield  {journal} {\bibinfo  {journal} {Phys. Rev. B}\
  }\textbf {\bibinfo {volume} {52}},\ \bibinfo {pages} {15122} (\bibinfo {year}
  {1995})}\BibitemShut {NoStop}%
\bibitem [{\citenamefont {Filipponi}(2000)}]{Filipponi2000}%
  \BibitemOpen
  \bibfield  {author} {\bibinfo {author} {\bibfnamefont {A.}~\bibnamefont
  {Filipponi}},\ }\href@noop {} {\bibfield  {journal} {\bibinfo  {journal} {J.
  Phys. B: At. Mol. Opt. Phys.}\ }\textbf {\bibinfo {volume} {33}},\ \bibinfo
  {pages} {2835} (\bibinfo {year} {2000})}\BibitemShut {NoStop}%
\end{thebibliography}%
 
\end{document}